\documentclass{article}

\usepackage{PRIMEarxiv}

\usepackage[utf8]{inputenc} 
\usepackage[T1]{fontenc}    
\usepackage{hyperref}       
\usepackage{url}            
\usepackage{booktabs}       
\usepackage{amsfonts}       
\usepackage{nicefrac}       
\usepackage{microtype}      
\usepackage{lipsum}
\usepackage{cite}
\usepackage{comment}
\usepackage{amsmath,amssymb,amsfonts}
\usepackage{algorithmic}
\usepackage{graphicx}
\usepackage{textcomp}
\usepackage{multirow}
\usepackage{makecell}
\usepackage[table,xcdraw]{xcolor}
\usepackage{bigdelim}
\usepackage{amssymb}
\usepackage[table]{xcolor}
\usepackage{fancyhdr}       
\usepackage{graphicx}       
\graphicspath{{media/}}     
\newcolumntype{?}{!{\vrule width 1.8pt}}

\title{A comparative study between paired and unpaired Image Quality Assessment in Low-Dose CT Denoising
}

\author{
  Francesco Di Feola \\
  Department of Radiation Sciences \\
  Umeå University \\
  Umeå, Sweden\\
  \texttt{francesco.feola@umu.se} \\
   \And
  Lorenzo Tronchin \\
  Research Unit of Computer Systems and Bioinformatics \\
  Campus Bio-Medico University of Rome \\
  Rome, Italy\\
  \texttt{l.tronchin@unicampus.it} \\
  \AND
  Paolo Soda \\
  Research Unit of Computer Systems and Bioinformatics \\
  Campus Bio-Medico University of Rome \\
  Rome, Italy\\
  Department of Radiation Sciences \\
  Umeå University \\
  Umeå, Sweden\\
  \texttt{p.soda@unicampus.it, paolo.soda@umu.se} \\
}

\begin{document}
\maketitle

\begin{abstract}
The current deep learning approaches for low-dose CT denoising can be divided into paired and unpaired methods.
The former involves the use of well-paired datasets, whilst the latter relaxes this constraint.
The large availability of unpaired datasets has raised the interest in deepening unpaired denoising strategies 
that, in turn, need for robust evaluation techniques going beyond the qualitative evaluation.
To this end, we can use quantitative image quality assessment scores that we divided into two categories, i.e., paired and unpaired measures. 
However, the interpretation of  unpaired   metrics is not straightforward, also because the consistency with paired metrics has not been fully investigated. 
To cope with this limitation, in this work we consider  15 paired and unpaired scores, which we applied to assess the performance of low-dose CT denoising.
We perform an in-depth statistical analysis  that not only studies the  correlation between paired and unpaired metrics but also within each category.
This brings out useful guidelines that can help researchers and practitioners select the right measure for their applications.
\end{abstract}

\keywords{Unpaired evaluation \and IQA \and generative adversarial network \and metrics}

\section{Introduction}
Medical images often contain noise and artifacts due to non-idealities of the imaging processes which hinder doctors’ ability to make an accurate diagnosis. 
Therefore, denoising is a critical and necessary step in the pre-processing stage of all medical imaging systems.
Among different imaging modalities, the use of Computed Tomography (CT), in clinical radiology, appears consolidated and reveals to be a crucial decision support system for medical diagnosis. Basing its functioning on the absorption of X-rays, it has been proved that ionizing radiations can cause damage at different levels in biologic material. Hence, with the aim of reducing and optimizing the radiation exposure under the ALARA (As Low As Reasonably Achievable) principle, the use of Low-Dose  acquisition protocol has become a clinical practice \cite{radiation_red} to be preferred over High-Dose protocols. However, if on the one side a reduction of radiations can be achieved, on the other side, the overall quality of the reconstructed CT decreases. Thereby, a trade-off between dose and noise level must be tackled, ensuring a sufficient diagnostic image quality. This is why several efforts have been dedicated towards searching denoising strategies, trying to obtain high quality CT images at the lowest cost in terms of radiation. \\ Denoising aims at making low dose CT (LDCT) images as close as possible to high dose CT (HDCT) images. The image corruption can be modeled as:
 \begin{equation}\label{eq:deg_process}
x = z(y) + \epsilon
\end{equation}
 where $y$ represents an HDCT image which is non linearly converted into a LDCT image $x$ by the noise model $z$ and an additive term $\epsilon$, which represents noise and other factors that cannot be modeled. The noise reduction task aims to fully recover $y$:
 \begin{equation}\label{eq:mapping}
\mathcal{M}: x \rightarrow \hat{y}
\end{equation}
 with $\mathcal{M}$ representing a generic denoising strategy and $\hat{y}$ the denoised image. Ideally, it should be $\hat{y}=y$ but in practice we have $\hat{y}\simeq y$.\\
 In the  literature, LDCT denoising is addressed either with \emph{paired} or \emph{unpaired} methodologies. The former  makes use of well-paired datasets, i.e., they need the low-dose images and their respective high dose counterparts, the latter do not need such a constraint alleviating the burden of collecting coupled images and fitting better the unpaired nature of the existing datasets \cite{wasser}.\\
Denoising can be evaluated through \emph{qualitative} and \emph{quantitative} image quality assessment (IQA) \cite{zhang2018can}.
While qualitative IQA relies on human experts' evaluation, quantitative IQA exploits mathematical functions  to objectively and reliably quantify the image quality, i.e., they measure the difference between $\hat{y}$ and $y$.
Quantitative IQA, depending on the nature of the dataset used, can perform paired or unpaired evaluations.  Again, the difference lies in the availability of paired datasets since, in this case, a reference image can be used to calculate the quality of a denoised image. This means that, given $x$, it is first denoised obtaining $\hat{y}$ and then compared with $y$.
To date, there is no gold standard of IQA for medical imaging due to various difficulties in designing suitable measurements able to capture heterogeneous characteristics and contents \cite{chow2016review} even though some authors have investigated the use of quantitative metrics in medical domains such as, for example, Magnetic Resonance Imaging \cite{zhang2018can, kastryulin2022image}.
This work experimentally evaluates quantitative IQA metrics, studying the correlation between paired and unpaired measurements to bring out the consistency of results between the less used unpaired metrics and the more widely used paired metrics, in the case of LDCT denoising.
\section{Materials}
We used a clinical dataset authorized by Mayo Clinic for 2016 NIH-AAPM-Mayo Clinic Low Dose CT Grand Challenge (\url{https://www.aapm.org/grandchallenge/lowdosect/}). It includes thoracic and abdominal HDCT and corresponding simulated LDCT data, using a quarter dose protocol. A total of 5376 CT slices from 9 patients were used in the training phase.
All the images were preprocessed as follows: 
first, all the raw DICOM files were converted in Hounsfield Unit (HU), second, we selected a display window centered in -500 HU with a width of 1400 HU, third, we normalized all the images in the range $[-1, 1]$ and eventually the images were resized to $256\times256$ resolution.
Even though the dataset is paired, meaning that a paired methodology could be used to approach the denoising task, all the training images were scrambled, in order to avoid paired correspondence between HDCT and LDCT images for each training batch. To evaluate the denoising performance,  we randomly extracted other 10 patients from the ``Low-dose CT image and projection data'' library by Mayo Clinic  available on The Cancer Imaging Archive  for a total of $3572$ images that underwent the same processing described above (\url{(https://www.cancerimagingarchive.net}). Code at:
https://github.com/FrancescoDiFeola/IQA-denosing

\section{Methods}
\begin{figure}[htbp]
    \centering
    \includegraphics[width=90mm]{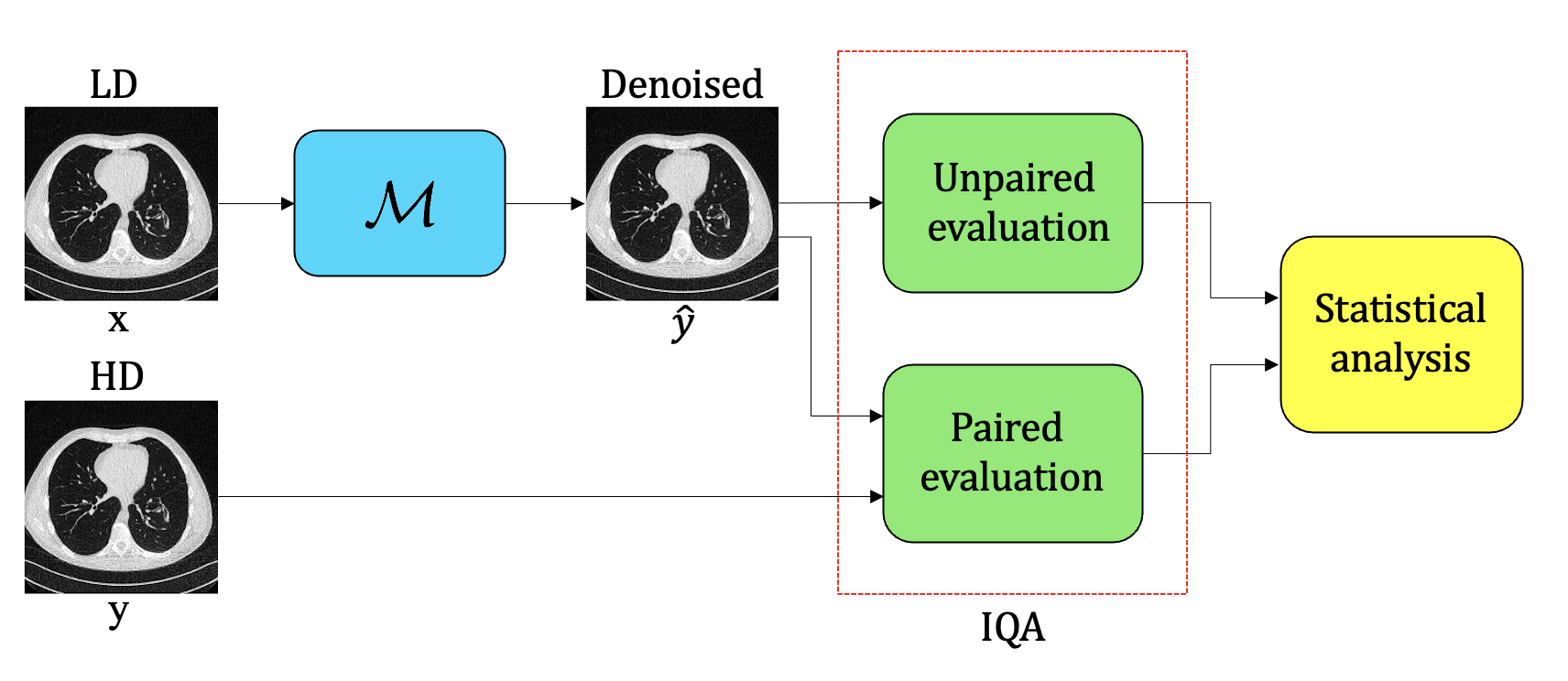}
    \caption{Pipeline of our methodology.}
    \label{fig:pipeline}
\end{figure}

The pipeline in \figurename~\ref{fig:pipeline} shows a schematic representation of this work highlighting some macro-blocks: denoising method ($\mathcal{M}$), IQA metrics (unpaired/paired evaluation) and statistical analysis which are now presented.

\subsection{Denoising method \texorpdfstring{$\mathcal{M}$}{M}}
\begin{figure}[htbp]
    \centering
    \includegraphics[width=75mm]{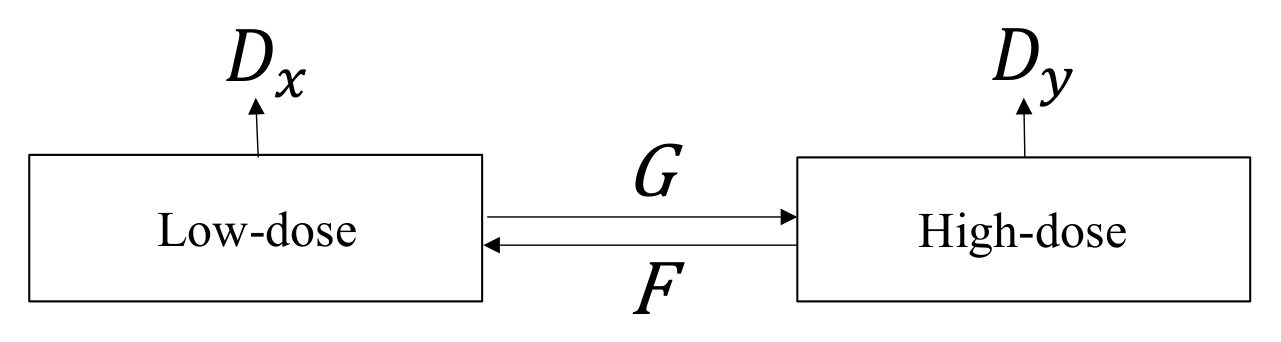}
    \caption{CycleGAN model performing the denoising task from low-dose to high-dose domain.}
    \label{fig:model}
\end{figure}

Denoising can be tackled with several approaches. The most investigated strategies go under the name of \emph{post-processing} methods, i.e.,
they apply denoising directly on the images after reconstruction. 
In this category, it is possible to include \emph{conventional methods} \cite{aharon2006k} which require prior knowledge or assumptions about the noise. 
To improve the generalization capacity of denoising algorithms, deep learning has shown promising performance \cite{li2020investigation} since they do not need strong assumptions and thus they can learn useful features while training, accommodating different kinds of data. This macro-category includes \emph{paired deep learning} methods using well-paired datasets; 
the algorithms can be CNN-based \cite{chen2017low1}, based on Residual Networks \cite{chen2017low2}, or employ the paradigm of adversarial training with Generative Adversarial Networks (GANs) \cite{wasser}.
The need for pairs represents a strong limitation for the actual application of these algorithms since the collection of coupled datasets is not only expensive and time-consuming, but it is also unfeasible from a clinical perspective \cite{li2020investigation, wasser}. 
Due to the intrinsic unpaired nature of the existing datasets, \emph{unpaired deep learning} methods seem to be the most suitable to address the problem of CT image denoising.
In this context, CycleGAN \cite{zhu2017unpaired} has the potential to overcome the limitations and difficulties of collecting paired datasets \cite{li2020investigation, wasser}. 
Its key aspect is that it guarantees an across-domain consistency called ``cycle-consistency''.
For image denoising, CycleGAN computes $\hat{y}$ by performing an unpaired image-to-image translation task, from a source domain (low-dose) to a target domain (high-dose), leveraging the use of generators ($G$, $F$) and discriminators ($D_{x}$, $D_{y}$), as shown in \figurename~\ref{fig:model}.

In this work, the CycleGAN framework was employed to perform denoising of low-dose CT images using a Tesla-V100-SXM2-32GB Graphic Processing Unit. We used the model proposed in \cite{zhu2017unpaired}, trained for $200$ epochs, with a batch size of $4$ and a learning rate equal to $10^{-4}$.

\begin{figure}[htbp]
    \centering
    \includegraphics[width=110mm]{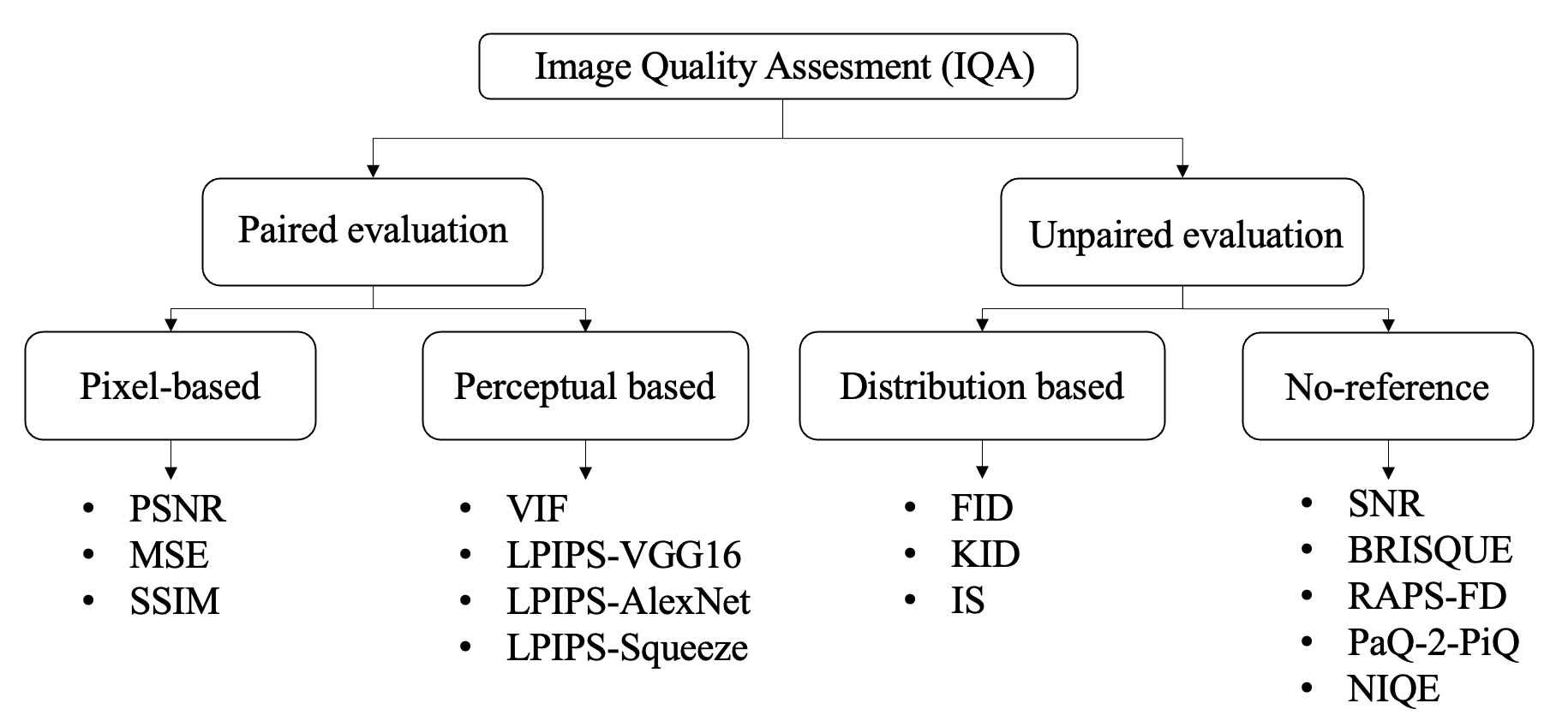}
    \caption{Proposed taxonomy of the IQA metrics.}
    \label{fig:taxonomy}
\end{figure}

\section{IQA metrics}
This section presents 15 IQA metrics examined in this work, which were select not only because  they are among the most used in the literature \cite{kastryulin2022image}, but also because they span different categories as shown in \figurename~\ref{fig:taxonomy}. 
This figure emphasizes that the evaluation methodology can be divided into \emph{paired} and \emph{unpaired} approaches: the former  needs pairs of images to be performed, whereas the latter relaxes this constrain.\\
Paired evaluation includes:
\begin{itemize}
    \item \emph{Pixel-based metrics}: they quantify an error-based distance between the denoised and the reference image.
    \item \emph{Perceptual-based metrics}: they quantify a perceptual or semantic based distance between the denoised and the reference image.
\end{itemize}
In the category of unpaired evaluation, it is possible to include:
\begin{itemize}
    \item \emph{distribution-based metrics}: they work quantifying a distance between distributions of images but without the need for paired correspondence.
    \item \emph{no-reference images}: they quantify the image quality directly on the images to be evaluated without any image reference.
    This category is more suitable for real applications where coupled images are not available. 
    However the interpretation of results is usually more challenging  due to its blind evaluation. 
\end{itemize}

Turning now the attention to the IQA metrics considered in this work and shown in \figurename~\ref{fig:taxonomy}, let us remember that $y$ is a reference HDCT image, whilst $\hat{y}$ is the denoised version computed from the LDCT image $x$. On these grounds, their formal definitions are:
\subsection{MSE (Mean-Squared-Error)}
It computes the absolute difference between intensities in pixels of the denoised and reference image:
\begin{equation}\label{eq:mse}
MSE(\hat{y},y) = \frac{1}{mn}\sum_{i=0}^{m-1}\sum_{j=0}^{n-1}\left[\hat{y}(i,j) - y(i,j)\right]^2
\end{equation}
where $(i,j)$ is the pixel coordinates, and $m$, $n$ are the numbers of columns and rows, respectively.
\subsection{PSNR (Peak-Signal-to-Noise-Ratio)}
 It compares the maximum intensity in the image ($MAX_{\hat{y}}$) with the error between $\hat{y}$ and $y$ given by the MSE \cite{wang2004image}:
\begin{equation}\label{eq:psnr}
PSNR(\hat{y},y)  = 10\times log_{10}\left(\frac{MAX_{\hat{y}}^2}{MSE(\hat{y},y)}\right)
\end{equation}
\subsection{SSIM (Structural Similarity Index)} It computes local similarity between two images as a function of luminance, contrast and structure \cite{wang2004image}:
\begin{equation}\label{eq:ssim}
SSIM(\hat{y},y) = \frac{(2\mu_{\hat{y}}\mu_{y} + C_{1})(2\sigma_{\hat{y}y} + C_{2})}{(\mu_{\hat{y}}^2 + \mu_{y}^2 +C_{1})(\sigma_{\hat{y}}^2 + \sigma_{y}^2 + C_2)}
\end{equation}
where $\mu_{\hat{y}}$, $\mu_{y}$ are mean intensities of pixels in $\hat{y}$ and $y$, respectively; similarly  $\sigma_{\hat{y}}^2$ and  $\sigma_{y}^2$ are the variances, $\sigma_{\hat{y}y}$ is the covariance whilst $C_{1}$ and $C_{2}$ are constant values to avoid numerical instabilities.
\subsection{VIF (Visual Information Fidelity)}
It is derived by a modeling of the Human-Visual-System \cite{sheikh2006image} in the wavelet domain:
\begin{equation}\label{eq:VIF}
VIF(\hat{y},y) = \frac{\sum_{j \in subbands}{I(y^{j}})}{\sum_{j \in subbands}{I(\hat{y}^j})}
\end{equation}
where $I(y^{j})$ and 
$I(\hat{y}^j)$  represent the information ideally extracted by the brain from a certain subband in the reference and test image, respectively. This score is consistent with subjective human evaluation and based on Natural Scene Statistics (NSS).
\subsection{LPIPS (Learned Perceptual Image Patch Similarity)}
It computes the similarity between the activations of two images for a specific network \cite{zhang2018unreasonable}. Let $\phi^{l}(\hat{y})$ and $\phi^{l}(y)$ denote the activation of the $l$th convolution layer of the network when processing $\hat{y}$ and $y$, respectively:
\begin{equation}
    LPIPS(\hat{y},y)= \sum_{l=1}^L \frac{1}{r c} MSE(\phi^{l}(\hat{y}), \phi^{l}(y))
\end{equation}
where $r \times c$ denotes the size of each feature map and $L$ is the number of the convolutional layers.
For completeness, we computed three LPIPS metrics using three different feature extractors $\phi$: VGG-16, Alex-Net and Squeeze-Net, which are named as LPIPS1, LPIPS2, LPIPS3, respectively.
\subsection{FID (Frèchet Inception Distance)}
It computes the Frechèt distance between the Inception feature vectors for denoised and reference images in the same domain \cite{burgos2022biomedical}.
\begin{equation}\label{eq:FID}
FID = |\beta_{y}-\beta_{\hat{y}}|+ tr(C_{y} + C_{\hat{y}} - 2\left(C_{y}C_{\hat{y}}\right)^{\frac{1}{2}})
\end{equation}
where $\beta_{y}$, $\beta_{\hat{y}}$ are feature-wise mean whilst $C_{y}$  $C_{\hat{y}}$ are covariance matrices.
\subsection{KID (Kernel Inception Distance)}
Similar to the FID, it computes the distance in terms of Maximum Mean Discrepancy (MMD) \cite{binkowski2018demystifying} between the distributions of denoised and reference images.
\subsection{IS (Inception Score)}
It takes into account diversity and quality of a distribution of denoised images. It is computed as the exponential of the expected value of the entropy, averaged over all denoised images \cite{burgos2022biomedical}.
\begin{equation}\label{eq:IS}
IS = exp(E_{x}KL(p(\hat{y}|x)||p(y)))
\end{equation} 
where $KL(p(\hat{y}|x)||p(y))$ is the Kullback-Leibler divergence between $p(\hat{y}|x)$ and $p(y)$, i.e., conditional and marginal distribution, respectively.
We compute the distributions involved in KID, FID and IS cropping each image into overlapping patches of size $50\times50$.
\subsection{SNR (Signal-to-Noise-Ratio)}
Widely used in clinic, it is defined as the ratio between the mean signal intensity measured in a tissue region of interest ($\mu_{signal}$)  and the standard deviation of the
signal intensity in an air region outside the object ($\sigma_{noise}$) \cite{zhang2018can}.
\begin{equation}\label{eq:SNR}
SNR = \frac{\mu_{signal}}{\sigma_{noise}}
\end{equation}
\subsection{BRISQUE (Blind/Referenceless Image Spatial Quality Evaluator)} It relies on the assumption that each distortion modifies NSS in an image.
It  extracts a feature vector from an image and then, using a support vector regression approach, it maps the feature vector to a numerical quality score \cite{mittal2012no}. 
\subsection{RAPS-FD (Radially Averaged Power Spectrum-Frèchet Distance)}
We introduce this new metric that  is computed in two steps. 
First, we calculate the   RAPS over $x$ and $\hat{y}$ \cite{aufrichtig2001measurement}; second, we compute  the Frechèt distance  to  quantify  the similarity between the two curves \cite{eiter1994computing}, considering location and ordering of the points along them.
\subsection{PaQ-2-PiQ (Patches Quality to Picture Quality)}
It is a blind quality predictor based on a modified ResNet architecture that leverages quality prediction on local patches to improve the prediction on the whole image \cite{ying2020patches}.
\subsection{NIQE (Natural image quality evaluator)} It bases its functioning on NSS features extracted from a corpus of natural images, fitting a multivariate gaussian model. The quality score on a test image computes the distance between the multivariate gaussian model fitted with NSS features extracted from the test image and the one fed with the NSS features from natural images \cite{mittal2012making}.

\section{Experimental evaluation} \label{section4}

\subsection{Statistical analysis}   \label{section4.1}
The Pearson Linear Correlation Coefficient (PLCC) and the Spearman Rank Order Correlation Coefficient (SROCC) are used to perform statistical analysis. Both are commonly used to illustrate the similarity between two sets of data \cite{chow2016review}.\\
PLCC ranges in $[-1,1]$, where the former and the latter indicates   a perfect negative and positive linear relationship between the two sets of data, respectively.\\ 
SROCC takes values in the same range as PLCC but instead of measuring linear relationship, it can be used to quantify non-linear monotonic relationships.
Moreover, we considered the absolute values of both scores since the focus is on whether or not different IQA metrics are correlated, regardless of the direction of correlation.
We also considered three ranges of correlation: 0-0.2, 0.3-0.5, 0.6-0.7, 0.8-1, indicating poor, fair, moderate and strong correlation~\cite{taylor1990interpretation}, respectively. 

\subsection{Importance analysis}
Besides statistical considerations, we investigate if it is possible to predict the values of paired metrics using the values of unpaired ones.
This would have two implications if we succeed with the prediction task: first,  there would exist a data driven relationship between paired and unpaired IQA. 
Second, this would open the chance to have an estimate of a paired metric even in an unpaired scenario, which is a useful functionality since for humans  interpreting paired metrics is easier than interpreting unpaired ones.
 A decision tree regressor performs the regression task: among the different regression algorithms that could be used, we opted for this algorithm since it does not require any assumption on the model underlying the data, it is intrinsically explainable and it also provides a measure of importance for each feature. We performed this analysis on the 10 patients in the test set, running a 10-fold cross validation. In other words, for each fold, suppose to fix a paired metric, called $A$, we trained the regressor using a training dataset where each sample, i.e., an image of the training set, is represented by a feature vector containing all the unpaired scores for that sample, while the ground truth is the paired metric $A$.
 This procedure is repeated using each paired metric as ground truth. 
 To evaluate the regression, we computed the Normalized Root Mean Squared Error (NRMSE) because  it makes the scores obtained for different regression metrics comparable, independently from their range.
Moreover, exploiting the intrinsic explainability of the regression trees, we extracted the feature importance, i.e., a raking of the most important unpaired metrics used to predict the paired metrics.

\section{Results}
This section presents the experimental results of our comparison between paired and unpaired metrics. The section is divided into three subsections: the first compares unpaired metrics within them, whilst the second performs a similar analysis on paired scores, and the third compares unpaired and paired scores.
\subsection{Unpaired vs Unpaired}
\tablename~\ref{tab:t2} reports the values of correlation between unpaired evaluation metrics, highlighting those that are distribution based and those that are no-reference. Furthermore, the table is organized so that the PLCC correlation scores and the SROCC scores are in the lower and upper triangles, respectively. The BRISQUE emerges as the metric with the highest inter-class correlation, i.e., it correlates well with distribution based metrics;
and the NIQE ranks second. While distribution-based metrics show higher intra-class correlation, which can be motivated by the fact that they all rely on the Inception Network, no-reference metrics appear to be less correlated among them: this suggests the fact that no-reference metrics are able to capture different aspects of the image quality encouraging diversity in the evaluation process.

\begin{table}[htbp]
\caption{Correlation between unpaired metrics:  PLCC and the SROCC values are   in the  lower and upper  triangle, respectively.}
\centering
\resizebox{12cm}{!}{
\begin{tabular}{lllll?lllll}
&                                
& \multicolumn{3}{c?}{\textbf{Distribution-based}}
& \multicolumn{5}{c}{\textbf{No-reference}}\\
\cline{3-10} 
& \multicolumn{1}{c|}{}          
& \multicolumn{1}{c|}{\rotatebox{90}{FID}}  
& \multicolumn{1}{c|}{\rotatebox{90}{KID}}  
& \multicolumn{1}{c?}{\rotatebox{90}{IS}}  
& \multicolumn{1}{c|}{\rotatebox{90}{SNR}}  
& \multicolumn{1}{c|}{\rotatebox{90}{BRISQUE}} 
& \multicolumn{1}{c|}{\rotatebox{90}{RAPS-FD}} 
& \multicolumn{1}{c|}{\rotatebox{90}{PaQ-2-PiQ}} 
& \multicolumn{1}{c|}{\rotatebox{90}{NIQE}}\\ 
\cline{2-10} 
\multicolumn{1}{c|}{\multirow{3}{*}{\textbf{Distribution-based}}} 
& 
\multicolumn{1}{c|}{FID}       
& \multicolumn{1}{c|}{-}    
& \multicolumn{1}{c|}{0.1}  
& \multicolumn{1}{c?}{0.25}  
& \multicolumn{1}{c|}{0.11} 
& \multicolumn{1}{c|}{0.23}    
& \multicolumn{1}{c|}{0.043}    
& \multicolumn{1}{c|}{0.19}      
& \multicolumn{1}{c|}{0.33}\\ 
\cline{2-10} 
\multicolumn{1}{c|}{}
& \multicolumn{1}{c|}{KID}       
& \multicolumn{1}{c|}{0.4}  
& \multicolumn{1}{c|}{-}    
& \multicolumn{1}{c?}{0.2}   
& \multicolumn{1}{c|}{0.1}  
& \multicolumn{1}{c|}{0.37}    
& \multicolumn{1}{c|}{0.074}    
& \multicolumn{1}{c|}{0.0083}    
& \multicolumn{1}{c|}{0.27}\\ 
\cline{2-10} 
\multicolumn{1}{c|}{}                    
& \multicolumn{1}{c|}{IS}        
& \multicolumn{1}{c|}{0.4} 
& \multicolumn{1}{c|}{0.62} 
& \multicolumn{1}{c?}{-}    
& \multicolumn{1}{c|}{0.12} 
& \multicolumn{1}{c|}{0.071}   
& \multicolumn{1}{c|}{0.12}     
& \multicolumn{1}{c|}{0.26}      
& \multicolumn{1}{c|}{0.046}\\ 
\Xhline{4\arrayrulewidth}
\multicolumn{1}{c|}{\multirow{5}{*}{\textbf{No-reference}}} 
& \multicolumn{1}{c|}{SNR}       
& \multicolumn{1}{c|}{0.045} 
& \multicolumn{1}{c|}{0.044} 
& 0.053 
& \multicolumn{1}{c|}{-}    
& \multicolumn{1}{c|}{0.12}    
& \multicolumn{1}{c|}{0.32}     
& \multicolumn{1}{c|}{0.11}      
& \multicolumn{1}{c|}{0.057}\\ 
\cline{2-10} 
\multicolumn{1}{c|}{}                    
& \multicolumn{1}{c|}{BRISQUE}   
& \multicolumn{1}{c|}{0.46} 
& \multicolumn{1}{c|}{0.48} 
& \multicolumn{1}{c?}{0.49}  
& \multicolumn{1}{c|}{0.13} 
& \multicolumn{1}{c|}{-}       
& \multicolumn{1}{c|}{0.011}    
& \multicolumn{1}{c|}{0.19}      
& \multicolumn{1}{c|}{0.34}\\ 
\cline{2-10} 
\multicolumn{1}{c|}{}                    
& \multicolumn{1}{c|}{RAPS-FD}  
& \multicolumn{1}{c|}{0.052} 
& \multicolumn{1}{c|}{0.038} 
& \multicolumn{1}{c?}{0.15}  
& \multicolumn{1}{c|}{0.4} 
& \multicolumn{1}{c|}{0.13}    
& \multicolumn{1}{c|}{-}        
& \multicolumn{1}{c|}{0.13}      
& \multicolumn{1}{c|}{0.35}\\ 
\cline{2-10} 
\multicolumn{1}{c|}{}                    
& \multicolumn{1}{c|}{PaQ-2-PiQ} 
& \multicolumn{1}{c|}{0.19} 
& \multicolumn{1}{c|}{0.047}  
& \multicolumn{1}{c?}{0.31}  
& \multicolumn{1}{c|}{0.14} 
& \multicolumn{1}{c|}{0.02}    
& \multicolumn{1}{c|}{0.064}     
& \multicolumn{1}{c|}{-}         
& \multicolumn{1}{c|}{0.053}\\ 
\cline{2-10} 
\multicolumn{1}{c|}{-}                    
& \multicolumn{1}{c|}{NIQE}      
& \multicolumn{1}{c|}{0.31} 
& \multicolumn{1}{c|}{0.3} 
& \multicolumn{1}{c?}{0.17}  
& \multicolumn{1}{c|}{0.034} 
& \multicolumn{1}{c|}{0.43}    
& \multicolumn{1}{c|}{0.4}      
& \multicolumn{1}{c|}{0.032}       
& \multicolumn{1}{c|}{-}\\ 
\cline{2-10} 
\end{tabular}
}
\label{tab:t2}
\end{table}

\subsection{Paired vs Paired}
\tablename~\ref{tab:3} shows the correlation between paired evaluation metrics which includes pixel-based and perceptual-based values, and it is organized as \tablename~\ref{tab:t2}.  The correlation values within pixel-based metrics are large and always higher than 0.6, so that can be considered between moderate and strong correlation. This result can be fully expected between PSNR and MSE, since they are formally dependent (see equation \ref{eq:mse}, \ref{eq:psnr}). Focusing now on perceptual based metrics, LPIPS show high values of correlation between them which suggests that different CNNs extract similar representations. Considering the inter-class correlation, i.e., between pixel-based and perceptual-based, the use of a different feature extractor in LPIPS does not seem to impact significantly, except for LPIPS1 (that uses a VGG feature extractor) which reports slightly less inter-class correlation. Among paired evaluation metrics, VIF reports the lowest correlation values both intra-class and inter-class which can be interpreted as a different information content extracted by this metric.
Unexpectedly, the LPIPS which by definition should be more focused on the extraction of perceptual/semantic information from the images, result to be highly correlated with metrics such as PSNR and MSE which instead quantify an ``error-based'' distance between pixels.

\begin{table}[htbp]
\caption{Correlation between paired metrics:  PLCC and the SROCC values are  in the  lower and upper  triangle, respectively.}
\centering
\resizebox{10cm}{!}{
\begin{tabular}{ccccc?cccc}
&
&\multicolumn{3}{c?}{\textbf{Pixel}} 
&\multicolumn{4}{c|}{\textbf{Perceptual}}\\
\cline{3-9} 
&\multicolumn{1}{c|}{} 
&\multicolumn{1}{c|}{\rotatebox{90}{MSE}} 
&\multicolumn{1}{c|}{\rotatebox{90}{PSNR}} 
&\multicolumn{1}{c?}{\rotatebox{90}{SSIM}} 
&\multicolumn{1}{c|}{\rotatebox{90}{VIF}} 
&\multicolumn{1}{c|}{\rotatebox{90}{LPIPS1}} 
&\multicolumn{1}{c|}{\rotatebox{90}{LPIPS2}} 
&\multicolumn{1}{c|}{\rotatebox{90}{LPIPS3}}\\ 
\cline{2-9} 
\multicolumn{1}{c|}{\multirow{3}{*}{\textbf{Pixel}}} 
&\multicolumn{1}{c|}{MSE} 
&\multicolumn{1}{c|}{-} 
&\multicolumn{1}{c|}{1} 
&0.78 
&\multicolumn{1}{c|}{0.65} 
&\multicolumn{1}{c|}{0.73} 
&\multicolumn{1}{c|}{0.83} 
&\multicolumn{1}{c|}{0.8}\\ 
\cline{2-9} 
\multicolumn{1}{c|}{} 
&\multicolumn{1}{c|}{PSNR} 
&\multicolumn{1}{c|}{0.86} 
&\multicolumn{1}{c|}{-} 
&0.78 
&\multicolumn{1}{c|}{0.65} 
&\multicolumn{1}{c|}{0.73} 
&\multicolumn{1}{c|}{0.83} 
&\multicolumn{1}{c|}{0.8}\\ 
\cline{2-9} 
\multicolumn{1}{c|}{} 
&\multicolumn{1}{c|}{SSIM} 
&\multicolumn{1}{c|}{0.61} 
&\multicolumn{1}{c|}{0.74} 
&-
&\multicolumn{1}{c|}{0.73} 
&\multicolumn{1}{c|}{0.81} 
&\multicolumn{1}{c|}{0.79} 
&\multicolumn{1}{c|}{0.73}\\ 
\Xhline{4\arrayrulewidth}
\multicolumn{1}{c|}{\multirow{4}{*}{\textbf{Perceptual}}} 
&\multicolumn{1}{c|}{VIF} 
&\multicolumn{1}{c|}{0.47} 
&\multicolumn{1}{c|}{0.66} 
&0.71 
&\multicolumn{1}{c|}{-} 
&\multicolumn{1}{c|}{0.66}
&\multicolumn{1}{c|}{0.67} 
&\multicolumn{1}{c|}{0.61}\\ 
\cline{2-9} 
\multicolumn{1}{c|}{} 
&\multicolumn{1}{c|}{LPIPS1} 
&\multicolumn{1}{c|}{0.61} 
&\multicolumn{1}{c|}{0.75} 
&0.85 
&\multicolumn{1}{c|}{0.66}
&\multicolumn{1}{c|}{-} 
&\multicolumn{1}{c|}{0.91}
&\multicolumn{1}{c|}{0.94}\\ 
\cline{2-9} 
\multicolumn{1}{c|}{} 
&\multicolumn{1}{c|}{LPIPS2} 
&\multicolumn{1}{c|}{0.8} 
&\multicolumn{1}{c|}{0.86} 
&0.78 
&\multicolumn{1}{c|}{0.72} 
&\multicolumn{1}{c|}{0.89} 
&\multicolumn{1}{c|}{-} 
&\multicolumn{1}{c|}{0.91}\\ 
\cline{2-9} 
\multicolumn{1}{c|}{} 
&\multicolumn{1}{c|}{LPIPS3} 
&\multicolumn{1}{c|}{0.79} 
&\multicolumn{1}{c|}{0.84} 
&0.7 
&\multicolumn{1}{c|}{0.62} 
&\multicolumn{1}{c|}{0.9} 
&\multicolumn{1}{c|}{0.94} 
&\multicolumn{1}{c|}{-}\\ 
\cline{2-9} 
\end{tabular}
}
\label{tab:3}
\end{table}

\subsection{Unpaired vs Paired}
\begin{figure}[htbp]
    \centering
    \includegraphics[width=90mm]{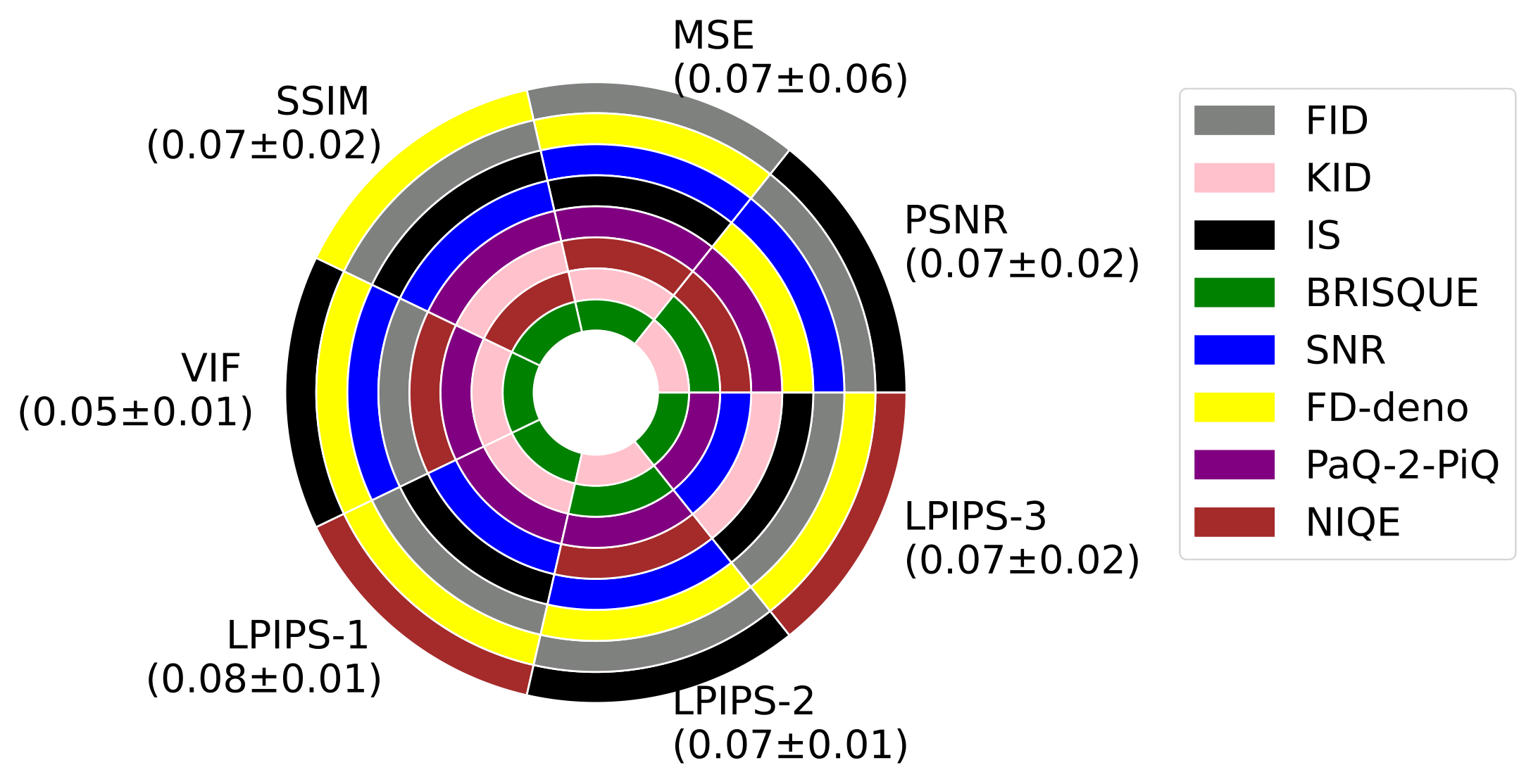}
    \caption{Importance analysis: each clove of the pie chart is a paired metric, and each portion of each circular crown corresponds to an unpaired metric used as a feature for the regression task. In round parenthesis, there is the value of the NRMSE.}
    \label{fig:pie2}
\end{figure}

 \tablename~\ref{tab:5} reports the average values of correlation of each unpaired metric with pixel-based, perceptual-based, and all paired values. In general, among the distribution-based, the highest score is obtained for the IS with PLCC correlation while in the no-reference category, the NIQE emerges in both types of correlation, followed by the BRISQUE with slightly lower values. 
 Overall, we found no strong difference in the correlation between unpaired and pixel/perceptual paired metrics. These findings confirm the correlation between pixel-based and perceptual-based metrics as discussed in the previous subsection.
 In this comparison, all the values lie in the range of poor or fair correlation which suggests  the difficulty of unpaired metrics to ensure high consistency with paired ones.\\
The results of the regression task are shown in \figurename~\ref{fig:pie2}, where the measure of importance for each feature is visualized through a pie chart.
Each clove of the pie chart corresponds to a different regression label, i.e., the paired metrics,  while each portion of the circular crown corresponds to a  different feature, i.e., the unpaired metrics. 
The outermost crown corresponds to the most important feature, while the  innermost crown corresponds to the less important one. 
Each feature is represented by a different color as reported in the legend.  
In \figurename~\ref{fig:pie2}, considering the ranking of the most important unpaired metrics used to predict the paired metrics, the IS results to be the most important feature to predict the values of PSNR, VIF and LPIPS-2.
The NIQE turns out to be the first feature in importance in LPIPS-1 and LPIPS-3 prediction. 
This result is consistent with the correlation values in which  the NIQE obtains the highest values of correlation both in PLCC and SROCC correlation (\tablename~\ref{tab:5}). 
Regarding distribution-based metrics, the IS assumes the highest values in PLCC correlation while this result is not confirmed by SROCC correlation in which the first position is disputed between the FID and the KID. The importance analysis helps at solving this uncertainty since it brings out that among distribution-based metrics the one which probably relates the most with unpaired metric is the IS, being the first most important feature in the prediction of PSNR, VIF and LPIPS-2.
Moreover, the importance analysis shows that the BRISQUE seems to be the least important in the prediction of paired metrics which is reasonable if we look at \tablename~\ref{tab:t2} where the BRISQUE reports high values of inter-correlation with distribution-based metrics and hence making a redundant contribution in the regression task. 
Another metric that seems to emerge from this analysis is the RAPS-FD which obtains the first position in predicting the SSIM while is second in predicting the MSE, the VIF, the LPIPS-1 and LPIPS-3. 
Finally, the values in round parenthesis in \figurename~\ref{fig:pie2} are the  NRMSE attained  in the inference phase of the regression task: it is worth noting that all such values are small and, hence, they suggest the robustness of the importance analysis described so far.

\begin{table}[htbp]
\caption{Average value of PLCC and SROCC (in round parenthesis) when comparing the unpaired IQA  vs the paired ones.}
\centering
\resizebox{13cm}{!}{
\begin{tabular}{c|c|c|c|c|}  
\cline{2-5}
 & & \bf Pixel-based & \bf Perceptual-based & \bf All Paired  \\
 \cline{2-5}
 &\bf FID &0.32$\pm$0.11 (0.17$\pm$0.02) & 0.37$\pm$0.09 (\textbf{0.19$\pm$0.06}) & 0.35$\pm$0.10 (0.18$\pm$0.05)\\
\cline{2-5}
 \bf Distribution &\bf KID & 0.34$\pm$0.11 (\textbf{0.25$\pm$0.09}) & 0.35$\pm$0.06 (0.18$\pm$0.03) & 0.34$\pm$0.09 (\textbf{0.21$\pm$0.07})\\
 \cline{2-5}
 \bf based&\bf IS & \textbf{0.36$\pm$0.14} (0.17 $\pm$ 0.07) & \textbf{0.38$\pm$0.12} (0.16$\pm$0.06)& \textbf{0.37$\pm$0.13} (0.17$\pm$0.06)\\
\Xhline{4\arrayrulewidth}
 &\bf SNR & 0.20$\pm$0.14 (0.17$\pm$0.07) & 0.12$\pm$0.11 (0.12$\pm$0.06) & 0.16$\pm$0.13 (0.18$\pm$0.10) \\
\cline{2-5}
 &\bf BRISQUE & 0.34$\pm$0.17 (0.22$\pm$0.15) & 0.32$\pm$0.12 (0.11$\pm$0.06) & 0.33$\pm$0.14 (0.16$\pm$0.12)\\
\cline{2-5}
\bf No-reference &\bf RAPS-FD & 0.31$\pm$0.17 (0.36$\pm$0.15) & 0.31$\pm$0.09 (0.39$\pm$0.09) & 0.31$\pm$0.13 (0.38$\pm$0.12) \\
 \cline{2-5}
 &\bf PaQ-2-PiQ & 0.07$\pm$0.04 (0.07$\pm$0.02) & 0.18$\pm$0.08 (0.17$\pm$0.07)& 0.13$\pm$0.09 (0.13$\pm$0.08)\\
\cline{2-5}
 &\bf NIQE & \textbf{0.41$\pm$0.06} (\textbf{0.41$\pm$0.07})  & \textbf{0.52$\pm$0.05} (\textbf{0.52$\pm$0.09}) & \textbf{0.47$\pm$0.08} (\textbf{0.48$\pm$0.10})\\
\cline{2-5}

\end{tabular}
}
\label{tab:5}
\end{table}

\begin{table}[h]
\caption{Average computational time per slice of the IQA metrics, using an Intel(R) Xeon(R) Gold 6226R CPU  2.90GHz.}
\centering
\resizebox{10cm}{!}{
\begin{tabular}{|c|c||c|c|}
\hline
\bf Paired metrics & \bf Average time (s) & \bf Unpaired metrics & \bf Average time (s)\\
\hline
\bf MSE & 0.000067$\pm$0.00001 & \bf FID & 0.2957$\pm$0.0130\\
\hline
\bf PSNR & 0.0001$\pm$0.000015 & \bf KID &  \bf 1.9347$\pm$0.4460 \\
\hline
\bf SSIM & 0.0037$\pm$0.0003 & \bf IS & 0.8381$\pm$0.1549 \\
\hline
\bf VIF & 0.2391$\pm$0.001 & \bf SNR & 0.0003$\pm$0.000018 \\
\hline
\bf LPIPS1 & 0.0686$\pm$0.0211 & \bf BRISQUE &  0.0188$\pm$0.0128\\
\hline
\bf LPIPS2 & 0.0112$\pm$0.0094 & \bf RAPS-FD  &0.01527$\pm$0.0094 \\
\hline
\bf LPIPS3 & 0.0223$\pm$0.0180 &  \bf PaQ-2-PiQ &0.115$\pm$0.0233\\
\hline
-&-&\bf NIQE &0.0191$\pm$0.0030\\
\hline
\end{tabular}
}
\label{tab:7}
\end{table}

\newpage
\subsection{Take-Home messages}
The above discussion brings out some useful considerations on  the metrics:
\begin{itemize}
    \item it is enough to use one pixel-based metric due to the high intra-class correlation;
    
    \item with reference to perceptual-based metrics we note that: i) the use of different CNNs as feature extractors in the LPIPS does not significantly affect the performance and the computational cost (\tablename~\ref{tab:7}); ii)  all the LPIPS scores report a high   correlation with pixel-based metrics,  which might indicate that their informative content is not so dissimilar the one extracted by pixel-based measures; iii) the VIF seems better suited to capture perceptual information;
    
    \item distribution-based metrics are redundant at least partially, and  might be reasonable to exclude the use of the KID due to its high computational cost (\tablename~\ref{tab:7});
    
    \item no-reference metrics report the lowest inter-class correlation values, which confirms the difficulty of this class of metrics in ensuring consistency of results with metrics that use a reference image.
    Nevertheless, the NIQE 
    might be the one  better approximating the evaluation performed by paired metrics;
    
     \item further to NIQE, in case of  unpaired evaluation we suggest selecting  one distribution-based metric among FID and IS;

     \item it is possible to predict to a larger extent the value of a paired metric using unpaired scores, a feature that could be beneficial when using unpaired data. 
\end{itemize}

\section{Conclusion} 
This work proposed a taxonomy of 15 IQA metrics dividing them into two groups based on the type of evaluation they can be used for, i.e., paired or unpaired evaluation. The analysis brought out that paired and unpaired evaluation lead generally to different results. However, the lack of consistency between these two types of evaluation might be interpreted as a complementary behavior that can work jointly. Nevertheless, we highlighted the most reasonable unpaired evaluation IQA metric that can be used in unpaired settings.

Further analysis should be carried out to improve the understanding of unpaired evaluation metrics with special attention to no-reference metrics. 
Indeed, since the latter have been designed for natural IQA, their effectiveness in medical imaging IQA should be further deepened with the possibility to optimize or to define novel metrics specific to this field.

\section{Acknowledgments}
The computations of this work were enabled by resources provided by the Swedish National Infrastructure for Computing (SNIC), partially funded by the Swedish Research Council through grant agreement no. 2018-05973.
We acknowledge financial support from: i) PNRR MUR project PE0000013-FAIR, ii) FONDO PER LA CRESCITA SOSTENIBILE Bando Accordo Innovazione DM 24/5/2017 (Ministero delle Imprese e del Made in Italy), under the project entitled ``Piattaforma per la Medicina di Precisione. Intelligenza Artificiale e Diagnostica Clinica Integrata'' (CUP B89J23000580005).

\bibliographystyle{unsrt}  

\bibliography{biblio.bib}

\end{document}